\documentclass[aps,prl,superscriptaddress,showpacs,twocolumn]{revtex4}
\usepackage{bm}
\usepackage{amsmath}
\usepackage{amssymb}
\usepackage{graphicx}
\usepackage{color}
\input{epsf}

\usepackage{caption}
\usepackage{subcaption}

\begin{document}

\title{Cooling of nanomechanical vibrations by Andreev injection}

\author{O.M.~Bahrova}
\affiliation{Center for Theoretical Physics of Complex Systems,
Institute for Basic Science, Daejeon, 34126, Republic of Korea}
\affiliation{B. Verkin Institute for Low Temperature Physics and
Engineering of the National Academy of Sciences of Ukraine, 47
Nauky Ave., Kharkiv 61103, Ukraine}

\author{S.I.~Kulinich}
\affiliation{B. Verkin Institute for Low Temperature Physics and
Engineering of the National Academy of Sciences of Ukraine, 47
Nauky Ave., Kharkiv 61103, Ukraine}

\author{L.Y.~Gorelik}
\affiliation{Department of Physics, Chalmers University of
Technology, SE-412 96 G{\" o}teborg, Sweden}

\author{R.I.~Shekhter}
\affiliation{Department of Physics, University of Gothenburg,
SE-412 96 G{\" o}teborg, Sweden}

\author{H.C.~Park}
\affiliation{Center for Theoretical Physics of Complex Systems,
Institute for Basic Science, Daejeon, 34126, Republic of Korea}

\date{\today}

\begin{abstract}
A nanoelectromechanical weak link composed of a
carbon nanotube suspended between two normal electrodes in a
gap between two superconducting leads is considered. The nanotube is treated as
a movable single-level quantum dot in which the position-dependent superconducting order parameter is induced due to the Cooper pair tunneling. We show that electron tunneling
processes significantly affect the state of the mechanical
subsystem. We found that at a given direction of the applied voltage
between the electrodes, the stationary state of the mechanical subsystem has a Boltzmann form with
an effective temperature depended on the parameters of the device.
As this takes place, the effective temperature
can reach significantly small values (cooling effect). We also
demonstrate that nanotube fluctuations strongly affect the dc
current through the system. The latter can be used to probe the predicted effects in an experiment.

\end{abstract}
\maketitle

\section{Introduction}

Nanoelectromechanical (NEM) systems promise to manipulate the mechanical motion of a nano-object using electronic dynamics~\cite{Ekinci, Cleland}. There are many approaches to control nanomechanical performance providing a number of new functionalities of nano-device operations, in particular, pumping or cooling of the mechanical subsystem~\cite{firstsh, belzig1,urgell,willick,zant}. One of the main approaches exploits the dc electronic flow through a nanosystem induced by either the bias voltage or temperature drop between two electronic reservoirs connected by a quantum dot (QD)~\cite{ilinskaya2, nazarov1,ilinskaya3,kulinich1,kulinich2}.

NEMS implementation hosts the nature of the coupling between the mechanical and electronic subsystems at nanoscale. It is  associated with localization of the electronic charge ~\cite{firstsh,nazarov1,Anton,anton2,belzig2,fedorets1} or spin~\cite{ilinskaya3,kulinich,atalaya} on the movable quantum dot. Nevertheless, the covalent coupling is a well-known concept in chemistry as a covalent bond based on sharing electron pairs between atoms and molecules. The incorporating of superconducting (SC) elements into NEMS allows one to use this coupling as a foundation for the electro-mechanical performance. A SC electrode located near a quantum dot can affect its electronic state via the tunneling exchange of Cooper pairs due to SC proximity effect. Additional injection of electrons from a biased normal metal electrode into the QD generates peculiar dynamics of the Cooper pairs on it. Such a process, which essentially involves Andreev conversion~\cite{andreev,kulik} of normal electrons into Cooper pairs, we have called in the title by Andreev injection. As a consequence, the interplay between coherent two-electron (Cooper pair) and incoherent single-electron tunneling into/out of the movable part of the NEMS may result in pumping or cooling effect~\cite{belzig1,belzig2}. Furthermore, if the tunneling amplitude depends on the distance between the QD and the SC leads, such exchange also provides a connection between the electronic and mechanical degrees of freedom.

In the  paper~\cite{arxiv1} a nanoelectromechanical weak link composed of the carbon nanotube suspended above a trench in a normal metal electrode and positioned in a gap between two superconducting leads, was considered. Such a setup is a generalization of the experimentally implemented one~\cite{AT}, where a CNT suspended between normal and superconducting electrodes. The nanotube has been treated as a movable single-level quantum dot, in which the position-dependent superconducting order parameter is induced as a result of Cooper pair tunneling. It has been shown that in such a system self-sustained bending vibrations can emerge if a constant bias voltage is applied between normal and superconducting electrodes. 

However, the semiclassical approach used in that paper does not allow the investigation of the operation of such a NEMS in the cooling regime. In this paper, using the reduced density matrix technique, we calculate the Wigner function characterizing a stationary state of the mechanical subsystem. We demonstrate that at certain conditions it has a Boltzmann distribution form with an effective temperature that can reach abnormally low values. The occurrence of this effect crucially depends on the direction of the bias voltage and the relative position of the quantum dot level. We also show that the nanotube fluctuations strongly affect the dc current through the system, a characteristic that can be used for direct experimental observation.

\section{Model}

\begin{figure}
\includegraphics[width=.9\columnwidth]{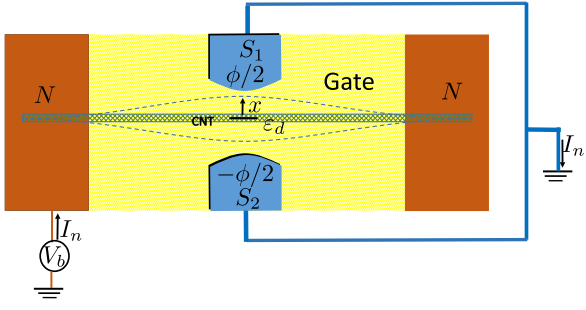}
\caption{Sketch of the nanoelectromechanical device under consideration. A carbon nanotube (CNT) is suspended in a gap between two edges of a normal electrode ($N$) and tunnel-coupled to it. Also, the CNT oscillates in the $x$ direction between two superconducting leads ($S_{1,2}$). This process affects the values of the tunneling barriers between the QD and superconducting electrodes. The normal electrode is biased by voltage $V_b$.
}\label{fig:fig1}
\end{figure}

 A schematic illustration of the NEMS investigated in this paper is represented in Fig.~\ref{fig:fig1}. A single-wall carbon nanotube (CNT) is suspended between two normal electrodes (with the same chemical potential) biased by the constant voltage $V_b$. Two
superconducting electrodes with the superconducting phase
difference $\phi$ are placed near the middle of the CNT in
such a way that the bending of the nanotube moves the nanotube
closer to one electrode and further away from the other. The
distance between quantized electronic levels inside the nanotube
is supposed to be much greater than other energy parameters. This fact allows one to
consider the nanotube as a single level QD.

The Hamiltonian of the system has the form
\begin{equation}\label{H}
H=H_d+H_v+H_l+H_t,
\end{equation}
where the first term $H_d$ describes the single-level QD,
\begin{equation}\label{Hd}
H_d=\sum_\sigma\varepsilon_d d^\dag_\sigma d_\sigma.
\end{equation}
The operator $d_\sigma^\dag (d_\sigma)$ is the creation (annihilation)
operator of the electron with the spin projection
$\sigma=\uparrow,\downarrow$ on the dot. The Hamiltonian $\hat
H_v$,
\begin{equation}\label{0035}
\hat H_v=\frac{p^2}{2m}+\frac{m\omega^2 x^2}{2},
\end{equation}
describes the mechanical dynamic of the dot, $p$ and $x$ are the
canonical conjugated momentum and coordinate,
$\left[p,x\right]=-\imath\hbar; \,m, \omega$ are the mass and
eigenfrequency of the dot correspondingly.

The third term in Eq.~(\ref{H}), $H_l=H_l^n+H_l^s$, describes the normal and
superconducting leads, respectively,
\begin{eqnarray}\label{Hl}
&& H_l^n=\sum_{k\sigma }(\varepsilon_k-eV_b)a_{k \sigma}^\dag
a_{k\sigma},\\
&& H_l^s=\sum_{k
j\sigma}\left(\varepsilon_{k}c^\dag_{kj\sigma}c_{kj\sigma}-
\Delta_{s}(\text{e}^{\imath\phi_j }
c^\dag_{kj\uparrow}c^\dag_{-kj\downarrow}+\text{H.c.})\right).
\end{eqnarray}
Here $a^\dag_{k\sigma} (a_{k\sigma})$, and $c^\dag_{kj\sigma}
(c_{k\sigma})$ are creation (annihilation) operators of the
electron with quantum number $k$ and spin projection $\sigma$ in
the normal and superconducting $j=1,2$ leads, respectively. 
$\Delta_s\text{e}^{\imath\phi_j}$ is the superconducting order
parameter (in the $j$ electrode). Note that energies
$\varepsilon_d,\varepsilon_k$ are counted from the Fermi energy of
superconductors. In what follows we suppose $\phi_1=-\phi_2=\phi/2$.

The Hamiltonian $H_t=H_t^n+H_t^s$ represents tunneling of
electrons between the dot and the leads,
\begin{eqnarray}\label{Ht}
&& H_t^n=\sum_{k\sigma}t^n_0 (a_{k\sigma}^\dag d_\sigma +\text{H.c.}), \\
&&  H_t^s=\sum_{k j\sigma}t^s_{ j}(x) (c^\dag_{kj\sigma}d_\sigma
+\text{H.c.}).\label{Hts}
\end{eqnarray}
Here the superconducting tunneling amplitude
$t^s_{1(2)}(x)=t^s_0 \text{e}^{\mp (x+a)/2\lambda}$ is position dependent, where
$2\lambda$ is the characteristic tunneling length and $a$ is the
parameter of the asymmetry. For a typical CNT-based nanomechanical
resonator, $2\lambda\sim 0.5$ nm \cite{GatecontrolledPE}. In what
follows we will concentrate our attention on the symmetric case,
$a=0$.

\section{Density matrix approach}
The time evolution of the system density matrix $\hat\rho$ is
described by the Liouville-von Neumann equation. We use the reduced density matrix approximation according to which
the full density matrix of the system $\hat{\rho}$ is factorized to the tensor
product of the equilibrium density matrices of the normal and
superconducting leads, and the dot density matrix as
$\hat\rho=\hat\rho_n\otimes\hat\rho_s\otimes\hat\rho_d$. Note 
that the reduced density operator $\hat \rho_d$ acts on the Hilbert
space which can be presented as the tensor product of the vibrational space of the
oscillator and the electronic space of the single electron level
on the QD.

In this paper we consider the stationary state of the system in the deep subgap case $\Delta_s\gg|eV_b|
\gg \Delta_d , \Gamma_n$, where $\Delta_d=2\pi\nu_s|t^s_0|^2,
\Gamma_n=2\pi\nu_n|t^n_0|^2$ ($\nu_{s(n)}$ is a density of states
in the superconducting (normal) electrode). Using the standard
procedure, one can trace out the leads degree of freedom and obtain
the following equation for the reduced density matrix $\hat\rho_d$
\cite{novotny}, 
\begin{equation}\label{EDM}
-\imath\left[H_d^{\text{eff}}+H_v,\hat\rho_d \right]+\mathcal
L_n\{\hat\rho_d\}+\mathcal L_\gamma\{\hat\rho_d\}=0,
\end{equation}
where
\begin{eqnarray}\label{Hdeff}
&&H^{\text{eff}}_d=H_d+\Delta_d(x,\phi)d_{\downarrow}d_{\uparrow}
+\Delta_{d}^* (x,\phi) d_\uparrow^\dag d_\downarrow^\dag,\\
&&\Delta_d(x,\phi)= \Delta_d\cosh(x/\lambda+i\phi/2).
\end{eqnarray}

In Eq.~(\ref{Hdeff}) $\Delta_d(x,\phi)$ is the off-diagonal order
parameter induced by the superconducting proximity effect
\cite{rozhkov}. The Lindbladian term in Eq.~(\ref{EDM}),
$\mathcal L_n\{\hat\rho_d\}$, is induced by the incoherent
electron exchange between the normal lead and the QD. The latter in
the high bias voltage regime, $|eV_b|\gg \varepsilon_0,\hbar\omega, T$, takes the
form
\begin{equation}\label{Ln}
\mathcal L_n\{\hat\rho\}=\Gamma_n\sum\limits_\sigma
\begin{cases}
2 d^\dag_\sigma\hat\rho d_\sigma-
\left\{d_\sigma d_\sigma^\dag ,\hat\rho \right\} , & \kappa =+1; \\
2 d_\sigma\hat\rho d^\dag_\sigma-\left\{ d^\dag_\sigma d_\sigma
,\hat\rho \right\} , & \kappa=-1;
\end{cases}
\end{equation}
where $\kappa=\text{sgn}(eV_b)$.

In Eq.~(\ref{EDM}) we phenomenologically introduce the dissipation
term $\mathcal L_\gamma\{\hat\rho_d\}$~\cite{petruccione},
\begin{equation}\label{001}
\mathcal L_\gamma\{\hat\rho\}=-m\omega\gamma\left(n_B+1/2\right)
\left[x, \left[x,\hat\rho\right]\right]-\imath \left(\gamma/2\right) \left[x,
\left\{p,\hat\rho\right\}\right],
\end{equation}
where $ \gamma$ is the damping rate, $n_B$ is the Bose-Einstein
distribution function,
\begin{equation}\label{605}
 n_B=\frac{1}{e^{\hbar\omega/T}-1},
\end{equation}
and $T$ is a temperature of the thermodynamic environment.

\begin{figure}
    \centering
    \begin{subfigure}[t]{0.85\columnwidth}
        \centering
        \includegraphics[width=\columnwidth, keepaspectratio]
        {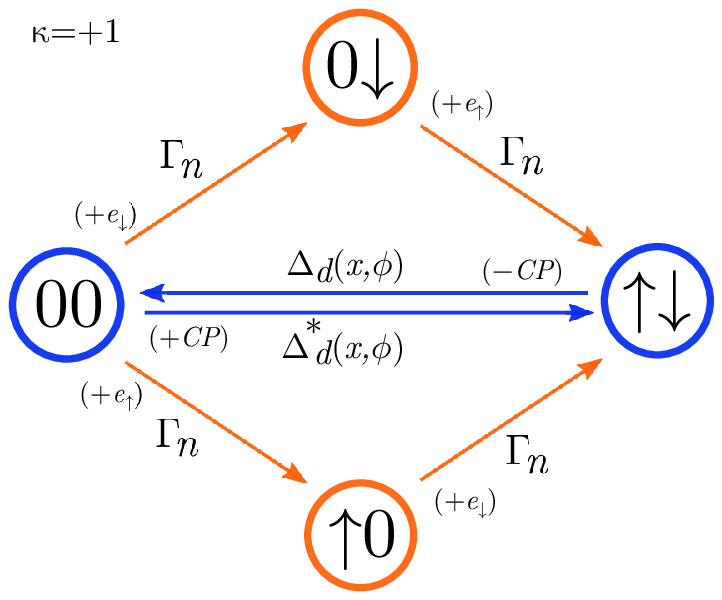}
        \caption{}\label{fig:subfig2a}
    \end{subfigure}
    \\
    \begin{subfigure}[t]{0.85\columnwidth}
        \centering
        \includegraphics[width=\columnwidth, keepaspectratio]
        {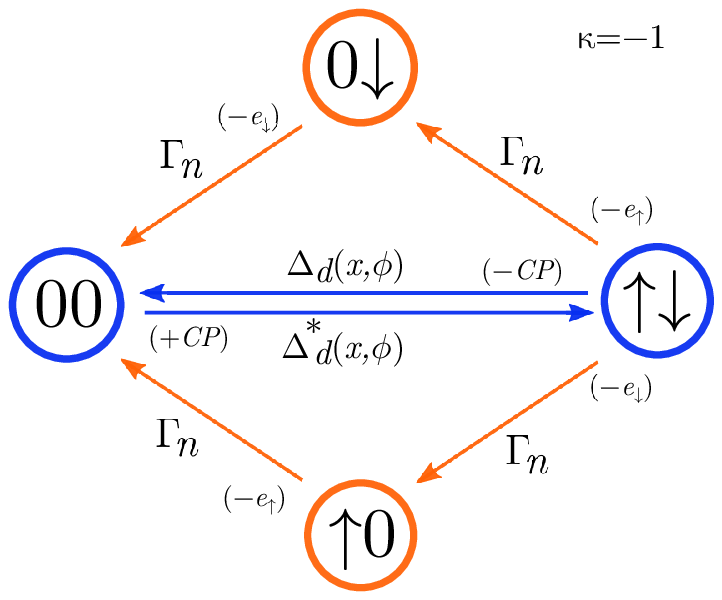}
        \caption{}\label{fig:subfig2b}
    \end{subfigure}
    \caption{Schematic illustration of the enabled transitions between electronic states in the quantum dot. The single-electron states change due to transitions from the empty to the single-occupied QD and then from the single-occupied to the double-occupied one (indicated by orange arrows). In the high bias voltage regime, the tunneling of electrons (a) or holes (b) with spin $\downarrow$ or $\uparrow$ is allowed only from the normal lead to the dot and forbidden in the opposite direction. Transitions between the empty and double-occupied QD are due to coupling with the superconducting leads (indicated by blue arrows). }\label{fig:fig2}
\end{figure}

Figure~\ref{fig:fig2} represents the electronic dynamics on the dot for two directions of the applied bias voltage,
$\kappa=\pm1$. Because of the considered parameter scales, not all electron processes are allowed. In the subgap regime, single-electron transitions between the dot and the superconducting leads are prohibited, and thus only an exchange of Cooper pairs occurs. Additionally, single-electron tunneling between the dot and the normal lead is enabled exclusively in one direction (from the lead to the dot, Fig.~\ref{fig:subfig2a}, or otherwise, Fig.~\ref{fig:subfig2b}) in the high bias voltage regime.

The state of the mechanical subsystem is completely
described by the reduced density matrix $\hat\rho_v=\text{Tr}
\hat\rho_d$, where the tracing operation is taken over the
electronic degrees of freedom on the dot. It is obvious that in the
 limiting case $\lambda\rightarrow\infty$ the electronic and
vibronic subsystems are independent and the reduced vibronic
density matrix has a form of equilibrium density matrix with
the effective temperature that is determined by an environment
temperature $T$. An alternative (and more visual) is the description
in terms of the Wigner distribution function
\begin{equation}\label{15}
W_v (x,p)=\frac{1}{2\pi}\int d\xi e^{-\imath p\xi}\left\langle
x+\frac{\xi}{2}\vert \hat \rho_v\vert x-\frac{\xi}{2}\right\rangle
\end{equation}
(we use the dimensionless variables: $x/x_0\rightarrow x,
px_0/\hbar\rightarrow p$, where $x_0$ is the amplitude of zero-point
oscillations, all energy parameters are measured in units of
$\hbar \omega$, the tunneling length $\lambda$ is measured in
units of $x_0, \gamma/\omega\rightarrow \gamma$).

The problem, Eqs.~(\ref{EDM})-(\ref{001}), can be solved  by the
perturbation expansion for the Wigner distribution function
$W_v(x,p)$,
\begin{equation}\label{A18}
W_v(x,p)\rangle=W_v^{(0)}(x,p)+ W_v^{(1)}(x,p) +...,
\end{equation}
using the smallness of the parameter $1/\lambda\simeq
10^{-2}-10^{-3}$~\cite{GatecontrolledPE} (or the parameter of electromechanical coupling,
$\Delta_d/\lambda\ll 1$). We have found (see Appendix for details of calculations) that the relevant
Wigner function which gives the probability distribution of the
vibrational amplitudes $A=\sqrt{x^2+p^2}$ in a stationary regime,
for $A\ll \lambda$ is a solution of the stationary Fokker-Planck equation,
\begin{equation}\label{204}
{\cal D}_1\frac{\partial}{\partial
A}\left(A^2W_v^{(0)}\right)+{\cal D}_2\frac{\partial}{\partial
A}\left(A\frac{\partial W_v^{(0)}}{\partial A}\right)=0.
\end{equation}
Here the drift, ${\cal D}_1$, and diffusive, ${\cal D}_2$,
coefficients take the form
\begin{eqnarray}\label{203}
&&{\cal D}_1= -\kappa\frac{\Delta_d^2\Gamma_n\varepsilon_d}
{\lambda^2 D_1}\sin^2(\phi/2)+\gamma,\\ \label{2040} &&{\cal
D}_2=\frac{\Delta_d^2\Gamma_n C}{\lambda^2
D_1}\sin^2(\phi/2)+\gamma\left(n_B+1/2\right),
\end{eqnarray}
where
\begin{eqnarray}\label{205}
&&\hspace{-0.55cm}D=\varepsilon_d^2+\Gamma_n^2+\Delta_d^2\cos^2(\phi/2),\\
&&\hspace{-0.55cm}D_1=\left(D-1/4\right)^2+\Gamma_n^2,\\
&&\hspace{-0.55cm}C=\frac{\left(D+1/4\right)\left(D+\varepsilon_d^2+
\Gamma_n^2 \right)-4\Delta_d^2\Gamma_n^2\cos^2(\phi/2)}{4D}.
\label{68}
\end{eqnarray}

The solution of Eq.~(\ref{204}) at small (in comparison to $\lambda$) values of the amplitude has
a form of the Boltzmann distribution function,
\begin{equation}\label{210}
W_v^{(0)}(x,p)=(\beta/\pi)\exp\left[-\beta\left(x^2+p^2\right)\right],
\end{equation}
where  the coefficient $\beta={\cal D}_1/2{\cal D}_2$.

The expressions, Eqs.~(\ref{203}), (\ref{2040}), define the
framework of validity of our consideration. It follows from Eqs.~(\ref{203})-(\ref{68}) that in the region which is related to the maximal cooling effect 
(the range of the values of parameters ($\phi, \varepsilon_d)$
near the point $\varepsilon_d=1/2, \phi =\pi$) the value of the level
width is restricted from below, $\Gamma_n\geq
\Gamma_n^{(0)}=\Delta_d^2/\lambda^2$.

\section{Ground-state cooling}

Nowadays, nanomechanical resonators with a significant value of the quality factor are achieved in experiments~\cite{Q5mil,Q2}. For such a case, the electromechanical coupling dominates the coupling with the thermodynamic environment, $1/\lambda\gg \gamma$. Thus, let us consider the case $\gamma\rightarrow 0$. From
Eqs.~(\ref{203})-(\ref{2040}) it follows  that the sign of the coefficient $\beta$ is determined by the sign of $\kappa\varepsilon_d$. If $\kappa\varepsilon_d$ is positive,
$\beta$ becomes negative. This situation corresponds to mechanical
instability of the system and it was discussed in Ref.~\cite{arxiv1}. In what
follows we consider the vibronic (stable) regime, when
$\kappa=-1,\,\varepsilon_d>0$ (the same for $\kappa=+1,\,\varepsilon_d<0$).

The coefficient $\beta$ determines the probability $P_0$ that the
system is in its ground state. In terms of Wigner distribution
functions this probability takes a form
\begin{equation}\label{56}
P_0=2\pi\int dx dp W_v^{(0)}(x,p)W_0(x,p)=\frac{2\beta}{\beta+1},
\end{equation}
where $W_0(x,p)=(1/\pi)\exp[-(x^2+p^2)]$ is the Wigner function of the harmonic oscillator
ground state. Note that according to Heisenberg's uncertainty
principle the maximal value of parameter $\beta$ is equal to
unity, $\beta<\beta_{\text{max}}=1$.

\begin{figure}
\centering
\includegraphics[width=0.85\columnwidth]{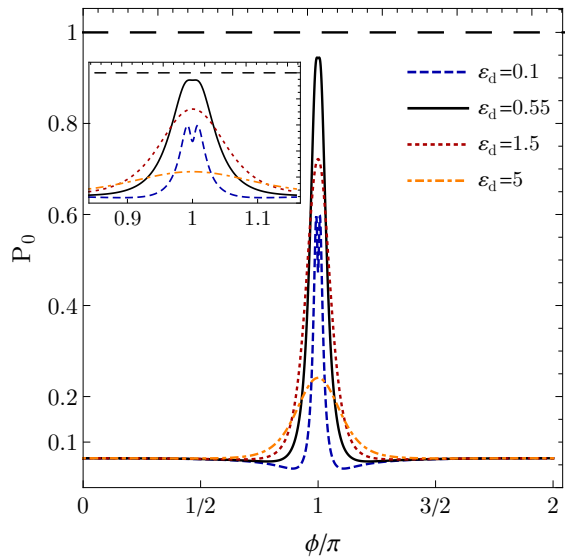}
\caption{The ground state occupation probability $P_0$ versus the
superconducting phase difference $\phi$ for different values of
the quantum dot energy level: $\varepsilon_d=0.1$ (blue
dashed curve), $0.56$ (black thick), $1.5$ (red dotted), $5$
(orange dot-dashed). The black dashed line indicates the maximal
value of the occupation probability. Inset: zoomed central region
where the cooling reaches its maximum at $\phi=\pi$. Other
parameters: $\Gamma_n=0.2; \Delta_d=25;\lambda=100; \gamma=
10^{-5}, T=15$.}\label{fig:fig3}
\end{figure}

Dependencies of the probability $P_0$ as
a function of the superconducting phase difference $\phi$ for
different values of the quantum dot energy level
$\varepsilon_d$ are demonstrated in Fig.~\ref{fig:fig3}

We see that the  maximal effect takes place in the "cooling region",
$\phi\simeq \pi, \varepsilon_d\simeq 1/2$, the degree of cooling
reaches the significant values, $P_0\simeq 0.95$. Note that the
maximal cooling occurs in the anti-adiabatic regime,
$\Gamma_n\simeq 0.2<1$.

\section{direct electric current}
The effects of cooling or heating of the mechanical vibrations can
be explored by dc current measurements. The Wigner distribution
function gives the possibility to calculate various physical
quantities.  The supercurrent in the $j$ superconducting lead is
determined by the change of the number of Cooper pairs and can be
presented as
\begin{equation}\label{currj}
I_j^{(s)}=\frac{2e}{\hbar}\text{Tr}\left(\frac{\partial
H_d^{\text{eff}}}{\partial \phi_j}\hat\rho_d\right).
\end{equation}
Due to the geometry of our system, the normal current is equal to
the sum of the partial currents corresponding to the superconducting
electrodes, $I_n=I_1^{(s)}+I_2^{(s)}$. In
terms of Wigner functions of the operators (see Appendix for details)
\begin{eqnarray}\label{W12}
&& \hat{\rho}_1=\text{Tr}\left[\left(d^\dag_\uparrow d^\dag_\downarrow+d_\downarrow d_\uparrow\right)\hat{\rho}_d\right],\\
&& \hat{\rho}_2=\imath\text{Tr}\left[\left(d^\dag_\uparrow d^\dag_\downarrow-d_\downarrow d_\uparrow\right)\hat{\rho}_d\right],
\end{eqnarray}
the expression for the current, Eq.~(\ref{currj}), takes a form
\begin{eqnarray}\label{curr}
&&I_n=e\omega\int dx dp
\left[\Delta_d\sin(\phi/2)\sinh(x/\lambda)W_1+\right.\nonumber\\
&&\hspace{2cm}+\left.\Delta_d
\cos(\phi/2)\cosh(x/\lambda)W_2\right].
\end{eqnarray}

The direct calculations of Eq.~(\ref{curr}) results in
\begin{equation}\label{currall}
I_n=I_0\frac{\Delta_d^2\cos^2{(\phi/2)}}{\Gamma_n^2+\varepsilon_d^2+
\Delta_d^2\cos^2(\phi/2)}+\mathcal{O}\left(1/\lambda^2\right),
\end{equation}
where $I_0=e\Gamma_n/\hbar$ (in dimension units). The leading term
in the expression, Eq.~(\ref{currall}), tends to zero in the limit
$\phi\rightarrow\pi$. Thus, at $\phi=\pi$ the current is determined
by the mechanical fluctuations and in the leading order of the
electromechanical coupling parameter it reads as
\begin{equation}\label{curr2}
I_n=I_0\left(\frac{\Delta_d}{\lambda}\right)^2
\frac{\left(\Gamma_n^2+\varepsilon_d^2+1/4\right)\langle
x^2\rangle+\varepsilon_d/2}
{\left(\Gamma_n^2+\varepsilon_d^2-1/4\right)^2+\Gamma_n^2},
\end{equation}
where the $\langle ...\rangle$ denote the
average value in the phase space with $W_v^{(0)}(x,p)$ and
$\langle x^2\rangle=(2\beta)^{-1}$.

\begin{figure}
\centering
\includegraphics[width=0.85\columnwidth]{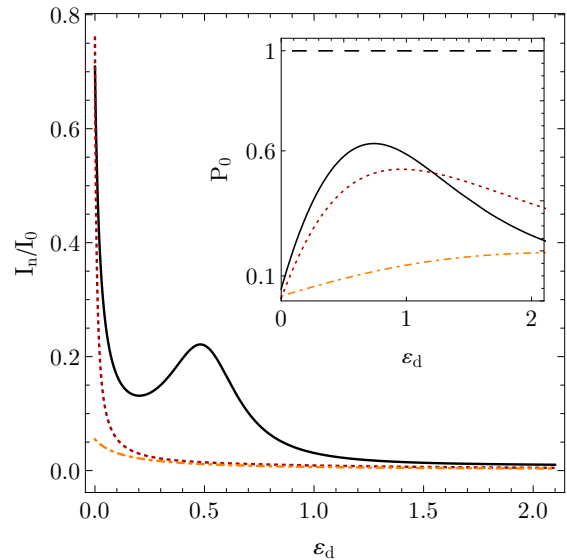}
\caption{The dependence of the electric current (normalized to
$I_0$) on the quantum dot level energy $\varepsilon_d$ at $\phi=\pi$ for
different values of $\Gamma_n:\,\Gamma_n=0.2$ (black
thick curve), $\Gamma_n=1$ (red dotted), $\Gamma_n=3$ (orange
dot-dashed). Inset: the ground state occupation probability versus
the QD level energy. The values $\Delta_d=5,\lambda=50, \gamma=5 \times
10^{-5}, T=15$ have been used.}\label{fig4}
\end{figure}

Figure~\ref{fig4} shows the  dependence of the electric current on the
quantum dot level energy $\varepsilon_d$ for different values of
$\Gamma_n$ at $\phi=\pi$.  We see that in the 
cooling regime  the  dependence of the electric
current has a pronounced minimum-maximum structure, that
disappears in the "heating" regime ($P_0\leq 0.5$). This fact can
serve as a criterion that the system is in the cooling regime.

\section{Conclusions}

We have considered the nanomechanical weak link that involves a carbon nanotube suspended between two normal leads and biased by a constant voltage. The nanotube, which is treated as a single-level quantum dot, performs bending vibrations in a gap between two superconducting electrodes. The coupling between the electronic and mechanical degrees of freedom is induced due to the superconducting proximity effect which exhibits in the appearance of the position-dependent dot order parameter. Using the density matrix approximation, we have found that at certain direction of the applied bias voltage, the stationary state of the mechanical subsystem has a Boltzmann form. Moreover, the probability to find the system in the ground state has been demonstrated to be $P_0\lesssim 1$. The latter is related to the cooling regime of the considered system. Additionally, the probability depends on the superconducting phase difference and the relative position of the QD energy level in a key manner. Also, we have discussed that the direct electric current behaviour mirrors the stationary state of the system. It can be served for an experimental detection of the predicted effects.

\section*{Acknowledgements}
O.M.B. thanks A.V.~Parafilo for helpful
discussions. Authors acknowledge the financial support from the IBS in Republic of Korea (IBS-R024-D1) and the NAS of
Ukraine (grant F 26-4) (S.I.K.).

\appendix

\setcounter{equation}{0}
\renewcommand\theequation{A.\arabic{equation}} 
\section {APPENDIX: EQUATIONS FOR THE WIGNER DISTRIBUTION FUNCTION}
The QD density matrix $\hat\rho_d$ acts in the Hilbert space that can be presented as a tensor product of the vibrational space of the harmonic oscillator and the Fock space of the single-level QD which is spanned
on the state vectors $\vert 0\rangle, d^\dag_\uparrow
(d^\dag_\downarrow )\vert 0\rangle =\vert \uparrow (\downarrow)
\rangle, d^\dag_\uparrow d^\dag_\downarrow\vert
0\rangle=\vert\uparrow \downarrow \rangle \equiv \vert 2 \rangle$.
We have got the following system of equations of
motion for electronic components of the
density matrix, $\hat{\rho}_d$ ($\kappa=+1$),
\begin{widetext}
\begin{eqnarray}\label{002}
&&\partial_t\rho_0=-\imath\left[H_v,\rho_0\right]-4\Gamma_n\rho_0-\imath
\Delta_d(x,\phi)\rho_{20}+\imath\rho_{02}\Delta_d^\ast(x,\phi)
+\mathcal L_\gamma\{\hat\rho_0\},\\
&&\partial_t\rho_\uparrow=-\imath\left[H_v,\rho_\uparrow\right]
+2\Gamma_n(\rho_0-\rho_\uparrow)+ \mathcal L_\gamma\{\hat\rho_\uparrow\},\\
&&\partial_t\rho_\downarrow=-\imath\left[H_v,\rho_\downarrow\right]
+2\Gamma_n(\rho_0-\rho_\downarrow)+\mathcal L_\gamma \{\hat
\rho_\downarrow\},\\
&&\partial_t\rho_{02}=
-\imath\left[H_v,\rho_{02}\right]+2\imath\varepsilon_d
\rho_{02}-2\Gamma_n\rho_{02}-\imath
\Delta_d(x,\phi)\rho_2+\imath\rho_0\Delta_d(x,\phi)+\mathcal
L_\gamma\{\hat\rho_{02}\},\\
&&\partial_t\rho_{20}=
-\imath\left[H_v,\rho_{20}\right]-2\imath\varepsilon_d
\rho_{02}-2\Gamma_n\rho_{20}-\imath
\Delta_d^\ast(x,\phi)\rho_2+\imath\rho_2\Delta_d^\ast(x,\phi)
+\mathcal L_\gamma\{\hat\rho_{20}\},\\
&&\partial_t\rho_2=-\imath\left[H_v,\rho_2\right]+2\Gamma_n
(\rho_\uparrow+ \rho_\downarrow)+\imath
\rho_{20}\Delta_d(x,\phi)-\imath\Delta_d^\ast(x,\phi)\rho_{02}
+\mathcal L_\gamma\{\rho_2\}.\label{003}
\end{eqnarray}

\end{widetext}
To find the equations in case of the opposite direction of the bias voltage, $\kappa=-1$, one needs to switch $0\rightleftarrows 2.$
The consequent  analysis of system, Eqs.~(\ref{002})-(\ref{003}), is
that to use the Wigner representation in the oscillator space,~Eq.(\ref{15}). We are interested in a steady state regime of the
mechanical subsystem in the limit when the parameter $1/\lambda$
is small. To find the solution of Eqs.~(\ref{002})-(\ref{003}) to
leading order in this parameter, it is convenient to introduce the
linear combinations of the Wigner distribution functions as
follows,
\begin{eqnarray}
&&W_v=W_0+W_\uparrow+W_\downarrow+W_2,\nonumber\\ 
&&W_0=W_0+W_2,W_1=W_{20}+W_{02}, \nonumber\\
&&W_2=\imath(W_{02}-W_{20}), W_3=W_0-W_2.
\end{eqnarray}\label{005}

In addition, it is convenient to change from ($x,p$) to polar
coordinates ($A,\varphi$) so that $x-\bar x=A \sin \varphi$ and $p=A\cos\varphi$,
where $\bar x\sim (1/\lambda)$ is an equilibrium
position of the dot. In the polar coordinates, the steady state
equation for the Wigner distribution function that describes the
mechanical degree of freedom, $W_v(A,\varphi)$, is given by the
equation (up to terms of the second order in the parameter
$1/\lambda$),
\begin{eqnarray}\label{009}
&&-\frac{\partial W_v}{\partial
\varphi}+ \bar x\hat T W_v +\gamma\left(n_B+1/2\right)\hat
T^2W_v\nonumber\\
&&\hspace{0.5cm}-\frac{\Delta_d}{\lambda}\sin(\phi/2)\hat T W_2+\frac{\Delta_d A}{\lambda^2}\cos(\phi/2)
\sin\varphi\hat T W_1\nonumber\\
&&\hspace{0.5cm}+\gamma\left(W_v+A\cos\varphi\hat T W_v\right)=0.
\end{eqnarray}
In Eq.~(\ref{009}) the differential operator $\hat T$ is defined
according to the expression,
\begin{equation}\label{A121}
\hat T=\cos\varphi\frac{\partial}{\partial A}-\frac{\sin\varphi}
{A}\frac{ \partial}{\partial \varphi}.
\end{equation}

Eq.~(\ref{009}) for the Wigner function
$W_v(A,\varphi)$ is coupled to the steady state equation for the
vector-function $\overrightarrow W=(W_0, W_1,W_2,W_3)^T$ that takes the following form (up to terms of
the first order in the parameter $1/\lambda$),
\begin{eqnarray}\label{010}
&&\hspace{2cm}-\frac{\partial \overrightarrow
W}{\partial\varphi}+2\hat M \overrightarrow W= \vec{F},\\
\label{A17}
&&\label{011} \hat M=\left(
\begin{array}{cccc}
-2\Gamma_n& 0 & 0 &- \kappa\Gamma_n\\
0 &-\Gamma_n&\varepsilon_d & 0 \\
0 &-\varepsilon_d&-\Gamma_n&-\Delta_d\cos(\phi/2)\\
0&0&\Delta_d\cos(\phi/2)&-\Gamma_n
\end{array}\right),\nonumber\\
&&\hspace{1cm}\label{012}\vec{F}=-\bar x \hat T \overrightarrow W
-2\Gamma_n W_v\left(\begin{array}{c} 1 \\0\\0 \\-\kappa\\
\end{array}\right)\nonumber+\\
&&\hspace{2.cm}+\frac{\Delta_d}{\lambda}\sin(\phi/2)
\left(\begin{array}{c} \hat T W_2\\2A\sin\varphi W_3\\\hat T
W_0\\-2A\sin \varphi W_1\\
\end{array}\right).\nonumber
\end{eqnarray}
Eqs.~(\ref{009})-(\ref{012}) have to be solved subject to the
periodic boundary conditions, $W_v(A,\varphi+2\pi)=
W_v(A,\varphi)$, $\overrightarrow W(A,\varphi+2\pi)=\overrightarrow
W(A,\varphi)$.

We solve these equations by perturbation expansions,
\begin{equation}\label{A181}
W_i(A,\varphi)\rangle=W_i^{(0)}(A,\varphi)+ W_i^{(1)}(A,\varphi)
+...,
\end{equation}
($i=v,0,1,2,3$), where $W_i^{(n)}$ is of
$n$:th order in $1/\lambda$.

It is evident from Eqs.~(\ref{009})-(\ref{A17}) that the functions
$W_v^{(0)}(A,\varphi), \overrightarrow W^{(0)}(A,\varphi)$ do not
depend on $\varphi$. Hence, $W_v^{(0)}(A,\varphi)=W_v^{(0)} (A)$
and
\begin{eqnarray}\label{A19}
&& W_0^{(0)}=\frac{\varepsilon_d^2+\Gamma_n^2+
(\Delta_d^2/2)\cos(\phi/2)}{D}W_v^{(0)},\\
&& W_1^{(0)}=\kappa\frac{\Delta_d\varepsilon_d\cos(\phi/2)}{D}W_v^{(0)},\\
&& W_2^{(0)}= \kappa\frac{\Delta_d\Gamma_n\cos(\phi/2)}{D}W_v^{(0)},\\
&& W_3^{(0)}=-\kappa\frac{\varepsilon_d^2+\Gamma_n^2}{D}W_v^{(0)},
\end{eqnarray}
where the coefficient $D$ is defined by Eq.~(\ref{205}).

From the requirement, $W_v^{(1)}(A,\varphi)=W_v^{(1)}(A)$, to
first order in the perturbation theory, Eq.~(\ref{009}) determines the
equilibrium position of the dot,
\begin{equation}\label{026}
\bar x=\kappa\frac{\Delta_d^2}{\lambda D}\sin(\phi/2)\cos(\phi/2).
\end{equation}
To second order in perturbation theory, Eq.~(\ref{009}) after
averaging over $\varphi$ variable takes the form,
\begin{widetext}
\begin{eqnarray}\label{20}
-\frac{\Delta_0\sin(\phi/2)}{\lambda A} \frac{\partial}{\partial
A} \left( A\left\langle\cos\varphi W_2^{(1)} \right\rangle\right)+\frac{\gamma}
{2A}\frac{\partial}{\partial A}\left(A^2W_v^{(0)}\right)
+\frac{\gamma\left(n_B+1/2\right) }{2A} \frac{\partial}{\partial A}\left(A
\frac{\partial W_v^{(0)}} {\partial A}\right)=0.
\end{eqnarray}
\end{widetext}

Here the brackets, $\langle f(A,\varphi)\rangle$, in Eq.~(\ref{20})
denote the zeroth Fourier component of the $2\pi$-periodic
function $f(A,\varphi)$; when deriving Eq.~(\ref{20}) we used the
property,
\begin{equation}\label{202}
\langle\hat T f(A,
\varphi)\rangle=\frac{1}{A}\frac{\partial}{\partial
A}\left(A\langle\cos\varphi f(A,\varphi)\rangle\right).
\end{equation}
Therefore, to get a closed equation for $W_v^{(0)}(A)$, one needs
to know the function $W_2^{(1)}(A,\varphi)$. To
first order in perturbation theory, this function can be determined from
Eqs.~(\ref{010})-(\ref{012}). As a result, one gets the stationary
Fokker-Planck equation for the oscillator Wigner distribution
function $W_v^{(0)}(A)$, Eq.~(\ref{204}).

\end{document}